\documentclass[twocolumn,showpacs,amsfonts]{revtex4-1}
\usepackage{amsmath}
\usepackage{latexsym}
\usepackage{float}
\usepackage{amssymb}
\usepackage{graphicx}
\usepackage{textcomp}
\usepackage{hyperref}
\usepackage{geometry}

 1

\def\ba{\begin{eqnarray}}
\def\ea{\end{eqnarray}}
\def\be{\begin{equation}}
\def\ee{\end{equation}}

\def\bm{\begin{math}}
\def\me{\end{math}}

\newcommand{\dummy}

\begin{document}
\title{Kinetics of Vapor-Solid Phase Transitions: Structure, growth and mechanism}
\author{Jiarul Midya and Subir K. Das$^{*}$}
\affiliation{Theoretical Sciences Unit, Jawaharlal Nehru Centre for Advanced Scientific Research,
 Jakkur P.O., Bangalore 560064, India}

\date{\today}

\begin{abstract}
~Kinetics of separation between the low and high density phases in a single component Lennard-Jones model 
has been studied via molecular dynamics simulations, at a very low temperature, in the space dimension $d=2$. 
For densities close to the vapor (low density) branch of the coexistence curve, disconnected clusters of 
the high density phase exhibit ballistic motion, the kinetic energy distribution of the clusters being
closely Maxwellian. Starting from nearly circular shapes, at the time of 
nucleation, these clusters grow via sticky collisions, gaining filament-like nonequilibrium 
structure at late times, with a very low fractal dimensionality. The origin of the latter is shown 
to lie in the low mobility of the constituent particles, in the corresponding cluster reference frame, 
due to the (quasi-long-range) crystalline order. Standard self-similarity in 
the domain pattern, typically observed in kinetics of phase transitions, is found to be absent 
in this growth process. This invalidates the common method, that provides a growth law same 
as in immiscible solid mixtures, of quantifying growth. 
An appropriate alternative approach, involving the fractality in the structure, 
quantifies the growth of the characteristic ``length'' to be a power-law with time,
the exponent being surprisingly high. The observed growth law has been 
derived via a nonequilibrium kinetic theory.
\end{abstract} 


\maketitle
\section{Introduction}
~Universality in dynamics of phase transitions is not as robust as 
statics \cite{AOnuki}. E.g. 
each of solid-solid \cite{KBinder1,AJBray,RALJones,SPuri1,IMLifshitz,SPuri2,SMajumder1}, 
fluid-fluid \cite{AJBray,SPuri1,KBinder2,KBinder3,EDSiggia,HFurukawa1,HFurukawa2,HTanaka1,HTanaka2,
HKabrede,SMajumder2,SRoy1,SRoy2,SMajumder3,SRoy3,KBinder4,GLeptoukh,RShimizu,FPerrot} and
solid-fluid \cite{GLeptoukh,MSleutel,YSuzuki} transitions may have different relaxation mechanism. 
In kinetics, the average size ($\ell$) of domains, rich in
one or the other type of particles, grows with time ($t$) as \cite{AJBray} $\sim t^{\alpha}$.
Domain growth in solid mixtures \cite{IMLifshitz} occurs via particle diffusion, providing $\alpha=1/3$.
In fluids \cite{AOnuki, AJBray}, hydrodynamics is important and the related mechanism 
depends upon the region of quench inside the miscibility gap 
\cite{AJBray,KBinder2,KBinder3,EDSiggia,HFurukawa1,HFurukawa2,HTanaka1,HTanaka2,SMajumder2,
SRoy1,SRoy2,SMajumder3,SRoy3}.
For disconnected spherical clusters, close to a coexistence curve, 
the fluid-fluid phase separations proceed via the cluster diffusion and coalescence 
mechanism \cite{KBinder2,KBinder3,EDSiggia}, for which $\alpha=1/d$ in space dimension $d$.
In the fluid-solid case, interplay between hydrodynamic
and particle-diffusion mechanisms can provide new growth laws.
Despite diverse relevance \cite{MSleutel,YSuzuki,GFCarnevale,GWetherill,SChandrasekhar,MHErnst}, 
other than condensed matter physics, understanding in this variety remains poor, due to 
difficulties with the identification of phases and slow nucleation.
\par
~Despite weak universality, self-similarity \cite{AJBray,KBinder2} is a robust phenomenon 
exhibited by structures during phase transitions. E.g. the above mentioned spherical clusters retain
their shape at all times. As a consequence, in usual scenario,
the two-point equal-time correlation function \cite{AJBray}
$C(r=|\vec{r}|,t)$ $(=<\psi(\vec{0},t)\psi(\vec{r},t)>-<\psi(\vec{r},t)>^{2}$, $\psi$ being a time and
space $(\vec{r})$ dependent order-parameter) obeys the scaling property \cite{AJBray,KBinder2}
$C(r,t)\equiv C(r/\ell(t))$, meaning, structures at two times differ only by size.
Even though found to hold \cite{SPuri2,SMajumder1,KBinder2,SRoy1,SRoy2,SRoy3,SKDas1}
in diverse situations, validity of this scaling remains unknown for the fluid-solid case. 
A crucial general question in this context is related to the consequence on growth if 
this ``simple'' scaling is absent.
\par
~Here we study vapor-solid transition in a Lennard-Jones (LJ)
model via hydrodynamics preserving molecular dynamics (MD) simulations. 
While the results are expected to have genral validity in $d=3$, here we focus on $d=2$,
relevant in contexts like active matter \cite{RWittkowski} and
bio-membranes \cite{MSleutel}. Strikingly, for quenches with low overall density,
the above scaling property is absent, having significant influence on the kinetics. 
This we understood to be due to the existence of two well-separated time 
scales in the problem, arising from the rigidity in the solid domains and
their fast hydrodynamic transport through the fluid phase, providing exotic fractal domains.
Such a rigidity is despite the facts that the background phase is very soft and a
long-range crystalline order cannot exist in $d=2$, for temperature $T>0$ and short-range interaction,
as stated in the Mermin-Wagner theorem \cite{NGoldenfeld}.
A new growth law emerges, mechanism of which has been identified that was hitherto unknown 
for phase transitions but has similarity with early stage clustering in a cosmic soup
\cite{GFCarnevale,GWetherill,SChandrasekhar,MHErnst}. The growth law has been
derived via a nonequilibrium kinetic theory \cite{GFCarnevale}.
\section{Model and Methods}
~For the kinetics, we perform MD simulations using a model \cite{SMajumder1,SRoy1,MPAllen} where
particles, at $r$ distance apart, interact via $U(r)=u(r)-u(r_c)-(r-r_c)(du/dr)_{r=r_c}$, $u(r)$ being 
the standard LJ form \cite{MPAllen} $u(r)=4\varepsilon [(\sigma/r)^{12}-(\sigma/r)^6)]$.
Here, $\varepsilon$ has the dimension of energy, $\sigma$ is the interatomic diameter, 
and $r_c$ ($=2.5\sigma$) is a cut-off distance. In $d=2$, the vapor-liquid 
phase diagram for this model was estimated by us via the Monte Carlo \cite{DPLandau} 
simulations in a Gibbs ensemble, the critical values of $T$ and number density ($\rho$) 
being respectively $T_c=(0.415\pm0.005)\epsilon/k_B$ and $\rho_c=0.350\pm0.005$, $k_B$ being 
the Boltzmann constant. We explore very low $T$ and $\rho$, viz., $0.25 \epsilon/k_B$ and $0.03$, 
to access the vapor-solid coexistence with disconnected morphology. The MD simulations 
were performed in a canonical ensemble, with the Lowe-Andersen 
\cite{EAKoopman}, dissipative particle dynamics \cite{SDStoyanov} 
and the Nos\'{e}-Hoover thermostats \cite{SNose,DFrankel} (NHT), which preserve hydrodynamics. 
We, however, have presented results only from the NHT that appears to us a better temperature controller. 
All our simulations were performed with the periodic boundary conditions, in square 
boxes of linear dimension $L=2048\sigma$. The quantitative results are presented after 
averaging over multiple initial configurations. The time in our MD simulations was measured in units of 
$(m\sigma^2/\varepsilon)^{1/2}$, $m$ being the the mass of a particle. 
For all presentation perposes we set $m$, $\sigma$ and $\varepsilon$ to unity. For the calculation of $C(r)$ we have
mapped our system \cite{SRoy1,SRoy2,SRoy3} onto a binary one by
assigning the values $\pm1$ to $\psi$ depending upon whether the local density is higher or
lower than the critical value.
\section{Results}
\begin{figure}
\centering
\includegraphics*[width=0.42\textwidth]{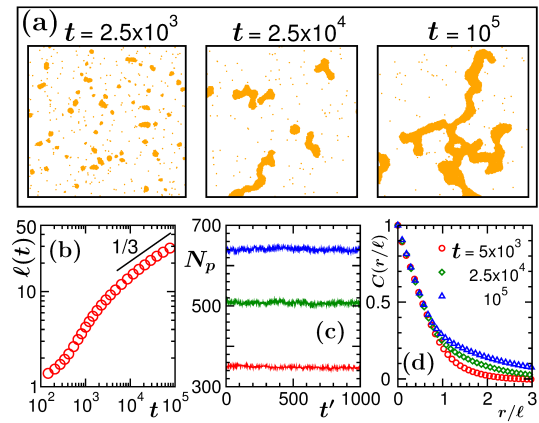}
\caption{\label{fig1} (a) Snapshots during the evolution of the Lennard-Jones system, 
having been quenched from a high temperature homogeneous state to the state point 
($\rho=0.03$, $T=0.25$) inside (and close to the vapor branch of) the coexistence curve. 
Only $400\times 400$ parts of the original systems ($L=2048$) are shown from three different times. 
(b) Average domain size, $\ell(t)$, obtained from the the length
distribution $P(\ell_d, t)$ as  $\ell(t)=\int P(\ell_d,t)~\ell_d ~ d\ell_d$,
a standard method \cite{SMajumder1,SRoy1,SRoy2,SRoy3} followed in
phase ordering dynamics, $\ell_d$ being the distance between two successive interfaces along any direction,
is plotted vs $t$, on log-log scale. The solid line corresponds to $ t^{1/3}$ behavior. 
(c) Plots of number of particles, $N_p$, 
vs translated time, for three different droplets. (d) Two point equal time correlation 
function from three different times are plotted vs $r/\ell$.}
\end{figure}

~Figure \ref{fig1}(a) shows some evolution snapshots 
(see Methods for parameter values and units). At earlier time, clusters are of nearly 
circular shape. With time the shape changes, the latest time pattern being fractal. 
Simulations with a stochastic (Andersen) thermostat \cite{DFrankel} provide 
circular pattern at all times. These fractal structures are then a result of hydrodynamics.
\par
~In Fig. \ref{fig1}(b) we show the $\ell$ vs $t$ plot. This quantity was calculated via a
standard method \cite{SMajumder1,SRoy1,SRoy2,SRoy3} followed in
phase ordering dynamics (see caption). In fluid-fluid phase separation,
for disconnected droplet morphology and growth via droplet-diffusion and coalescence mechanism, 
one writes \cite{KBinder2,KBinder3,EDSiggia}, for the droplet density $n$,
\begin{eqnarray}\label{bs_law}
dn/dt=-Bn^{2},
\end{eqnarray}
where $B$ ($=D\ell$, $D$ being a diffusivity) is a positive constant, decided by the 
Stokes-Einstein-Sutherland relation \cite{JPHansen}. Solution of equation (\ref{bs_law}), 
taking $n \propto 1/\ell^{d}$, provides $\alpha=1/d$. But the late time data appear consistent 
with $\alpha=1/3$ which points towards the particle-diffusion (from a smaller droplet to a larger one)
mechanism in solid mixtures \cite{AOnuki,KBinder1,RALJones,AJBray,SPuri1,IMLifshitz}. 
However, as seen in Fig. \ref{fig1}(c), the number of particles in a droplet, 
before it undergoes a collision, remains constant. Thus, the growth occurs via 
collisions. The consistency of the data with $\alpha=1/3$ is accidental. 
\par
~As seen in Fig. \ref{fig1}(d), there exists serious lack of scaling \cite{SKDas2} in the $C(r)$ which 
may in general be due to a disproportionate growth in the structure. 
E.g. in a fractal morphology the branches may have a different widening rate compared to the rate of 
increase in the overall length \cite{RLJack}. In the present case this can be due to 
two time scales, coming from different transport mechanisms. 
While the overall growth occurs due to motion of the clusters and collisions among them, 
the structural change of the clusters, following the collisions, is related to  
the dynamics of the constituent particles. The mobility of these particles, in their 
cluster reference frame, can be low, depending upon the phase of the clusters which 
will be addressed later. In such cases, scaling property of $C(r)$ involves the
fractal dimensionality and extraction of a characteristic length via the standard 
procedure is not meaningful. An alternative is to look for the growth of 
the average mass ($M$), appropriate \cite{SPaul} for non-percolating structures as in 
the present case, and connect it to the average radius of gyration \cite{TVicsek} ($R_g$). 
The latter is the true characteristic length scale, and is different from $\ell$ for the present problem.
\begin{figure}
\centering
\includegraphics*[width=0.45\textwidth]{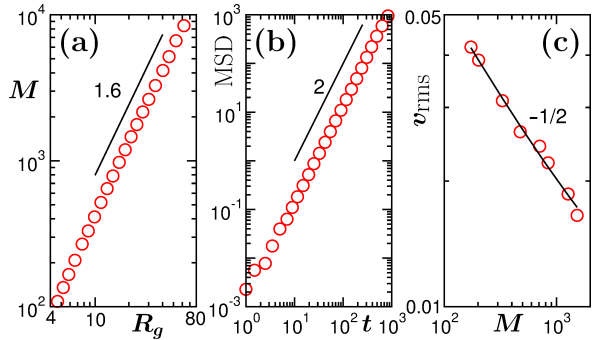}
\caption{\label{fig2}
(a) Log-log plot of average (time dependent) cluster mass $(M)$ vs the average radius of gyration ($R_g$). 
The solid line corresponds to a power-law behavior with the exponent $d_f=1.6$. 
(b) Mean-squared-displacement of a typical cluster is plotted, on a log-log scale, vs $t$. 
The $t^2$ line corresponds to ballistic motion. (c) Root-mean-squared 
velocity, $v_{\mbox{rms}}$, of the clusters is plotted, on log-log scale, vs $M$. 
The solid line has a power-law decay with exponent $1/2$.}
\end{figure}
\par
~In Fig. \ref{fig2}(a) we present a plot of $M$ vs $R_g$. This shows a power-law behavior 
\begin{eqnarray}\label{rg_m}
M \sim R^{d_f}_g. 
\end{eqnarray}
\begin{figure}
\centering
\includegraphics*[width=0.42\textwidth]{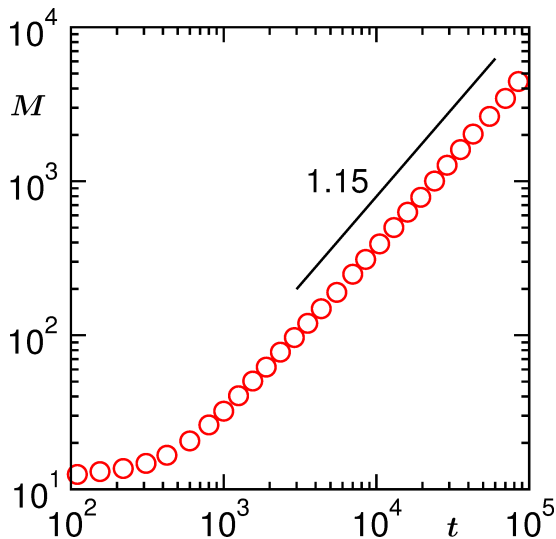}
\caption{\label{fig3}
Log-log plot of $M$ vs $t$. The solid line corresponds to a power-law growth with the exponent $1.15$.}
\end{figure}
The data appear consistent with $d_f=1.6$, the fractal or mass dimensionality \cite{TVicsek} 
of the structure. The dimensionality is low, even compared to the structures formed in diffusion 
limited aggregation \cite{TAWitten}. This we will explain later.
To apply equation (\ref{bs_law}), we need to know about the nature of motion of these fractal objects.
Note that despite the dominant mechanism being collision, if the droplet motion is not 
diffusive and the structure is not circular, $B$ need not be a constant. In Fig. \ref{fig2}(b), we show 
the mean-squared-displacement (MSD) \cite{JPHansen} 
($=\langle |{\vec R}_{\mbox{CM}}(t)-{\vec R}_{\mbox{CM}}(0)|^2 \rangle$) 
for the center of mass (CM) of a cluster (${\vec R}_{\mbox{CM}}$ being the position vector for the CM), 
vs time. The robust $t^2$ behavior confirms a ballistic motion. These findings 
necessitate the call for the theory of ballistic aggregation, having the equation 
\cite{GFCarnevale,SNPathak}
\begin{eqnarray}\label{dndt_baslistic}
\frac{dn}{dt}=-\mbox{``collision-cross-section''}~\times< v_{\mbox{rel}}>\times~n^2.
\end{eqnarray}
In $d=2$, the ``collision-cross-section'' is a ``length'', equaling 
$R_g$. For uncorrelated motion of the droplets \cite{GFCarnevale}, the mean 
relative velocity, $<v_{\mbox{rel}}>$, of the clusters should be the root-mean-squared velocity, 
$v_{\mbox{rms}}$, which, as seen in Fig. \ref{fig2}(c), varies with $M$ as 
\begin{equation}\label{vrms}
v_{\mbox{rms}}\sim M^{-1/2},
\end{equation}
expected for Maxwellian distribution of kinetic energy, validity of which is separately checked.
Incorporating equations (\ref{rg_m}) and (\ref{vrms}) in 
equation (\ref{dndt_baslistic}), along with $n \propto 1/M$, we obtain $dM/dt \sim M^{(2-d_f)/2d_f}$,
providing 
\begin{eqnarray}\label{m_solution}
 M \sim t^{2d_f/(3d_f-2)}.
\end{eqnarray}
The value of the exponent for $d_f=1.6$ is approximately $1.143$. In Fig. \ref{fig3} we show a plot 
of $M$, vs $t$. After a flat nucleation regime, the data exhibit a power-law behavior. 
The exponent is very much consistent with the predicted value.
This implies 
\begin{eqnarray}\label{rg_vs_t}
R_g \sim t^{0.72},
\end{eqnarray}
the exponent being rather high for a $2D$ phase separation.
Given the clean nature of the data we do not aim for a finite-size scaling
analysis \cite{DPLandau,MEFisher, SMajumder3}.
\begin{figure}
\centering
\includegraphics*[width=0.4\textwidth]{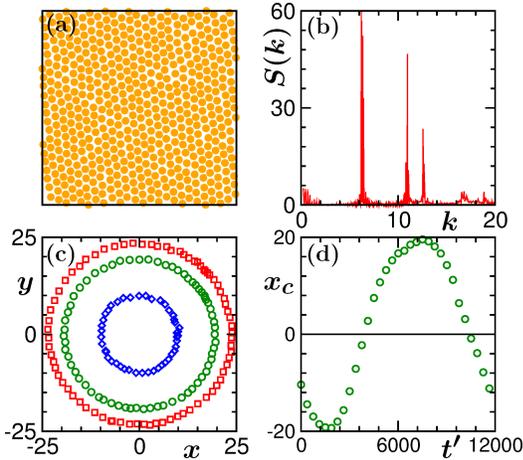}
\caption{\label{fig4}
(a) Snapshot of a part of a cluster at $t=10^5$.
(b) The structure factor $S(k)$
($=<\sum_{i=1,j=1}^{N}
\mbox{exp}(i\vec{k}\cdot\vec{r})>/N;~ \vec{r}=\vec{r}_i - \vec{r}_j$, $N$
being the number of particles in the considered cluster), 
for the snapshot in (a), is plotted vs the wave number $k$.
(c) Relative positions of three particles with respect to the center of mass of a cluster
to which these particles belong. (d) Plot of $x$-component 
for the intermediate circle in (c), vs translated time.}
\end{figure}
\par
~On the question of another (slower) time scale being the origin of fractality, we intend to
show the existence of a rigid crystalline order. In Fig. \ref{fig4}(a) we show the
snapshot of a part of a cluster and in Fig. \ref{fig4}(b) the corresponding structure
factor \cite{JPHansen}, $S(k)$, vs the wave number $k$. These clearly
confirm the presence of a crystalline order. Nevertheless, because of only quasi-long-range
order \cite{NGoldenfeld} in a soft background and the possibility that frequent particle exchanges
between the solid and the vapor phases can take place in the interfacial regions,
one should examine the rigidity to validate the solid state transport properties. 
In Fig. \ref{fig4}(c), we show trajectories of three particles, belonging to the same cluster, 
with respect to the cluster CM. All the scalar distances remain constant. 
The circular shape of the trajectories confirm rotation of the cluster. 
Such rotations make the collisions more probable in the peripheral regions, perpendicular to the major 
cluster-axes. This explains the very low value for the fractal dimension.
In Fig. \ref{fig4}(d), we show the $x$-component of the intermediate trajectory, 
vs time. It appears, the typical time period of 
rotations is comparable but smaller than the time scale of the collisions. This also confirms the 
rigidity of the clusters over a long time. Thus, the origin of the fractal domain structure 
lies in the rigid crystalline order. The constituent particles of a non-circular cluster cannot 
rearrange themselves to provide the latter a circular shape before it undergoes a collision.
\section{Conclusion}
~In conclusion, for overall density close to the vapor branch, 
the vapor-solid transition proceeds unusually fast, via the ballistic aggregation mechanism 
\cite{GFCarnevale}. The standard self-similarity, usually observed in phase transitions, is violated 
during the growth process, due to formation of filament-like fractal pattern.
Invoking the fractality, along with the mass dependence of the mean relative velocity 
of the clusters, in a general formalism \cite{GFCarnevale}, we have obtained the growth law.
The identified mechanism has similarity with that 
for the formation of clusters in cosmic dust and may as well find application in 
coarsening in active matters where the particles in a domain move coherently,
in addition to other experimental situations. We expect the results to have  
general validity, outside the dimension considered here.
\vskip 0.1cm
\textbf{Acknowledgment:} 
The authors are thankful to K. Binder, D. Dhar and J. Horbach for
critical comments. They acknowledge financial support from Department of Science and Tech-
nology, India, via Grant No SR/S2/RJN-13/2009, as well as from the Marie Curie Actions Plan
of European Commision (FP7-PEOPLE-2013-IRSES grant No. 612707, DIONICOS). JM is
grateful to UGC, India, for research fellowship.

\vskip 0.2cm
${*}$ das@jncasr.ac.in

\end{document}